\documentclass[12pt]{iopart}
\usepackage{iopams}

\usepackage{epsfig}
\usepackage{bm}

\begin{document}
\title[Discrete Gravity in the first order formalism] {Discrete Gravity as a local theory of the Poincar\'e 
group in the first order formalism}

\author{Gabriele Gionti S.J.{${}^{1,2}$}\footnote[3]{e-mail:ggionti@as.arizona.edu}}
\address{{${}^{1}$} Vatican Observatory Research Group,
Steward Observatory, 933 North Cherry Avenue, The University of Arizona, Tucson AZ 85721 U.S.A.}
\address{{${}^{2}$}Specola Vaticana, V-00120 Citt\'a Del Vaticano, Vatican City State}

\begin{abstract}
A discrete theory of gravity, locally invariant under the Poincar\'e
group, is considered as in a companion paper. We define a first order
theory, in the sense of Palatini, on the metric-dual Voronoi complex 
of a simplicial complex. We follow the same spirit of the continuum theory of
General Relativity in the Cartan formalism.
The field equations are carefully derived taking in account the constraints
of the theory. They look very similar to first order Einstein continuum equations
in the Cartan formalism. It is shown that in the limit of {\it small deficit
angles} these equations have Regge Calculus, locally, as the only solution. A quantum measure 
is easly defined which does not suffer the ambiguities of Regge Calculus, and a 
coupling with fermionic matter is easily introduced.   
\end{abstract}


\maketitle

\section{Introduction}
In spite of many recent developments, in particular in String Theory and Loop Quantum Gravity, 
the problem of quantization of General Relativity is still an open one both from mathematical and 
physical points of view. 
A key point of General Relativity is its close resemblance (and yet not equivalence) to an ordinary gauge 
theory. This is very clear in the first order (vierbein) formalism, where the dynamical variables are those
suitable for a gauge theory of the Poincar\'e group. This has been the starting point of the Loop Quantum Gravity 
approach too.
This similarity led to various attempts to write down a regularized version of GR on a lattice. One of the most 
well-known is Regge Calculus \cite{regge}, which is a natural 
and geometrically appealing way of discretizing gravity on a simplicial lattice. Recently it has been an active
field of research in connection with Dynamical Triangulations, two dimentional Quantum Gravity, 
and therefore with conformal field theory and
String Theory.
One of the weak points of the simplicial lattice is that it is naturally related 
to the second order (metric) formalism of GR, so that its connection with gauge theory is
rather obscure. The aim of this article is to continue the work started by M. Caselle, A. D'Adda and
L. Magnea \cite{Ale}, to reformulate Regge Calculus in terms of dynamical variables belonging to the Poincar\'e 
group, so that it makes an explicit connection with gauge theory in general and Wilson-like gauge theory on a  
lattice \cite{Wilson} in particular. We have also found strict analogies between this approach and discrete Ashtekar variables
\cite{immirzi}.
We pursue this goal starting in section II by summarizing the main results of \cite{Ale} 
in the second order formalism which 
consists of assigning the rotational degrees of freedom (Lorentz connection) to the links of a dual metric complex (Voronoi complex)
of the original simplicial complex, so that they are functions of the translational degrees of freedom. 
In section III and IV we formulate a first-order principle, in the sense of Palatini \footnote[1]{The 
labeling of the first order method as {\it Palatini} has been questioned by F.W. Hehl, {\it Four lectures on 
Poincar\'e gauge field theory}, Erice Cosmology Inst. 1979, pag. 40 footnote. See also Ferraris  
et al. {\it ...Palatini method discovered by Einstein 1925}, Gen. Relat. and Grav. 14, 
pag. 243 (1982)\label{pippo}.},
in which the independent variables are the Lorentz connection and 
the normals to the $n-1$-faces. These normals are considered analogues of the $n$-bein 
in the continuum theory, and the connection 
matrices as the connection one-form in GR. In section V we prove, in the case of {\it small deficit angles},
that Regge Calculus is a solution in the first-order formalism. This result is not obvious if we vary 
independently the two sets of the variables above. Moreover this equivalence for small deficit angles tells us
that this first-order formalism has the  effect of smoothing out some pathological
configurations, like {\it spikes}, which affect Regge Calculus and might prevent the theory from having a smooth
continuum limit. These {\it spike} configurations are in fact in the region of large deficit angles, where the first-order and  
second-order formalisms are not equivalent. In section VI we derive the general field equations for the connection matrices and for the 
$n$-bein. In section VII a measure for the path integral of this discrete theory of gravity
is introduced and is shown to be locally invariant under $SO(n)$. 
As a last step in section $VII$ we propose a coupling of this discrete theory of gravity with fermionic matter.
This coupling is entirely performed by following the general prescription of the continuum theory.
In other formulations of gravity (Regge Calculus and Dynamical Triangulations) the coupling with fermionic matter 
is usually introduced {\it ad hoc}.     
    
\section{Second-order formalism}
In 1961 T. Regge \cite{regge} (see also \cite{ruth} for a recent and updated summary 
of Regge Calculus and its alternative approaches)  proposed a discretized version of General 
Relativity now known as Regge calculus. It is, mainly, based on the idea 
of substituting a  Piecewise-Linear (PL)-manifold for differential manifold. In 
particular, these are simplicial manifolds built by gluing together two distinct 
$n$-dimensional flat simplexes by one and only one $n-1$-dimensional 
face. The final product of this construction is called a simplicial complex, 
which owns the manifold structure if it has been 
made in such a way that each point of the simplicial complex  has  a
neighbourhood homeomorphic to ${\Bbb R}^{n}$ \cite{seifert}. The simplicial manifolds, 
 we consider, are orientable. On each $n-2$-dimensional 
simplex $h$, called a {\it hinge}, a deficit angle $K(h)$ is defined

\begin{equation}
K(h)= 2\pi - \sum_{\sigma^{i}_{h} \supset h}\theta(\sigma^{i}_{h},h)
\label{apex}
\end{equation}

\noindent where $\sigma^{i}_{h}$, $i=1,...,p$ is one of the $p$ 
$n$-dimensional simplices incident on $h$, and
$\theta(\sigma_{h},h)$ its dihedral angles on $h$. 
The dihedral angle of a $n$-simplex on the hinge 
is the angle between the $n-1$-dimentional faces that have the hinge in common. 
$K(h)$ and the volume of the hinge $V(h)$ can be expressed \cite{hamber}  
as functions of the squared length of the one-dimensional simplices (edges) of
the complex.  
The Einstein-Regge action is, in analogy to the continuum case, 
a functional over
the simplicial manifolds and depends on the incidence matrix of the simplicial 
complex \cite{frohlich} and on the (squared) lengths of the edges. Usually, in 
Regge calculus, the incidence matrix is fixed, so that the action can be written as 

\begin{equation}
S_{R}=\sum_{h}K(h)V(h)
\label{reggeact}\;\;\;\; .
\end{equation}

\noindent The corresponding 
Einstein equations \cite{regge} are derived 
by requiring that the action is stationary 
under the variation of the length of the edges.

\noindent The original aim of this theory was to give approximate solutions of 
the Einstein equations in the case in which the topology is known. The theory is, as 
stressed by Regge himself, completely coordinate independent.

\noindent It was pointed out in reference \cite{Ale} that a theory of Regge Calculus locally invariant 
under the Poincar\'e group can be formulated 
by choosing in every simplex an orthonormal reference frame. 
In this 
way every $n$-dimensional simplex, considered as a piece of ${\Bbb R}^{n}$, can be 
seen as local inertial reference frame, being flat, and an n-bein base can be 
chosen in it. As in the continuum theory, we need a connection between 
the reference frames of two simplexes that share a common $n-1$-dimensional simplex.

\begin{figure}

\epsfxsize=8.0truecm
\epsfysize=8.0truecm
\centerline{\hbox{\epsffile{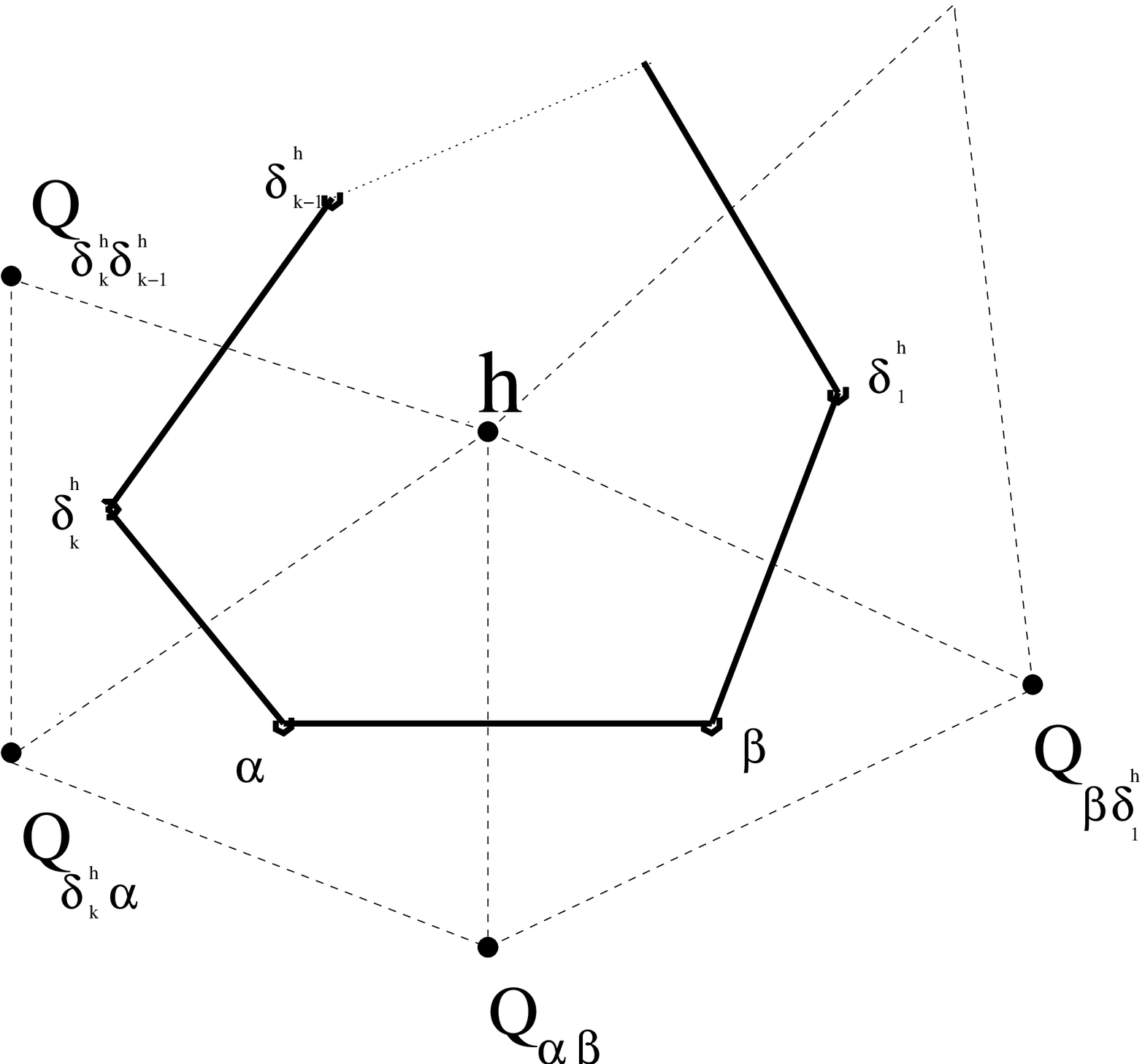}}}

\caption{\label{giorno1} Dual Voronoi Plaquette}

\end{figure}

\noindent Now we summarize and update  some definitions and results of \cite{Ale} that 
will be useful for our future discussions.
Consider an hinge $h$ and let $\{P_1^{h},...,P_{n-1}^{h}\}$ be its vertices. 
Suppose that this 
hinge is shared by $k+2$ $n$-simplices (see FIG.1) 
$\{\alpha,\beta,\delta^{h}_{1},...,\delta^{h}_{k}\}$ whose vertices are labelled in this way 

\begin{eqnarray}
\alpha&\equiv& \{P^{h}_1,...,P^{h}_{n-1},Q_{\delta^{h}_{k}\alpha},Q_{\alpha\beta}\}\nonumber\\ 
\beta&\equiv& \{P^{h}_1,...,P^{h}_{n-1},Q_{\alpha\beta},Q_{\beta\delta^{1}_{h}}\}\nonumber\\
.....&\equiv&.....................................................................\nonumber\\
\delta^{h}_{k}&\equiv& \{P^{h}_1,...,P^{h}_{n-1},Q_{\delta^{h}_{k}\delta^{h}_{k-1}},Q_{\beta^{h}_{k}\beta}\}\;\;\;\; 
\label{vertices}
\end{eqnarray}

\noindent In each simplex $S_{\alpha}$  we can choose an
origin and an orthogonal reference frame. In this frame the vertices of the simplex $S_{\alpha}$ have 
the following coordinates

\begin{eqnarray}
P_{i}&\equiv& \{y_{i}^{a}(\alpha)\} \;\;\; a=1,...,n\;,\; i=1,...,n-1 \nonumber\\
Q_{\delta^{h}_{k}\alpha}&\equiv& \{z^{a}_{\delta^{h}_{k}\alpha}(\alpha)\} \nonumber \\
Q_{\alpha\beta}&\equiv& \{z^{a}_{\alpha\beta}(\alpha)\}
\label{alpha}
\end{eqnarray}

\noindent Given a $n$-dimensional simplicial complex, there exists a general procedure for defining the 
dual metric complex, called the Voronoi dual \cite{Ale}
\cite{senechal} \cite{okabe} (see also \cite{thesis} for an easy treatment
and calculation at the details). The $n$-dimensional Voronoi polyhedron dual to a vertex $P$ 
of the simplicial complex is the set of the points of the simplicial complex closer to $P$ 
than to any other vertex, using the standard flat metric in the simplex. It turns out that the $k$-simplex
of the simplicial complex is dual to a $(n-k)$-polyhedron in the dual Voronoi complex , and the $k$-dimensional linear 
space identified by the $k$ simplex is orthogonal to the $(n-k)$ dimensional space spanned by the corresponding 
polyhedron. In particular the point dual to the $n$-simplex $S$ is the centre of a $(d-1)$-dimensional sphere 
passing through all vertices of $S$ (look at FIG.1).

\noindent We can associate uniquely to a dual Voronoi 
edge  an element 
of the Poincar\'e group $\Lambda(\alpha,\beta)\equiv 
\{ \Lambda^{a}_{b}(\alpha,\beta), \Lambda^{a}(\alpha,\beta) \}$ by requiring that, 

\begin{eqnarray}
\Lambda^{a}_{b}(\alpha,\beta)y^{b}_{i}(\beta)
+\Lambda^{a}(\alpha,\beta)&=&y^{b}_{i}(\alpha)\nonumber \\
\Lambda^{a}_{b}(\alpha,\beta)z^{b}_{\alpha\beta}(\beta)
+\Lambda^{a}(\alpha,\beta)&=&z^{b}_{\alpha\beta}(\alpha)\;\;\;\; .
\label{connec}
\end{eqnarray}

\noindent In other words we are embedding $\alpha$ and $\beta$ in ${\Bbb R}^{n}$ and adopting 
 the standard notion of parallel displacement in ${\Bbb R}^{n}$. 
 We move the origin of the reference frames of $\beta$ to $\alpha$. So 
 that the position vectors in $\beta$ of the vertices of the common face
  $S_{\alpha}\cap S_{\beta}$ will become coincident with 
  the position vectors of the same vertices in $\alpha$. 
 Thus the matrix 
 $\Lambda^{a}_{b}(\alpha,\beta)$ will be an orthogonal matrix which describes the change 
 from the reference frame of  $\beta$ to $\alpha$, 
 considered now as two different reference frames
 of the same vector space \cite{frohlich}. 
 Since the simplicial manifold is orientable we can choose the reference frame 
 in $\alpha$ and $\beta$ in such a way that $\Lambda^{a}_{b}(\alpha,\beta)$ is 
 a matrix of SO(n). This procedure 
 determines a unique connection that is the a torsion-free {\it Levi Civita or Regge connection} \cite{frohlich} .
 We can assign to each vertex $D_{\alpha}$, of the dual Voronoi complex, its coordinates $x^{a}({\alpha})$ in the frame
 of $\alpha$. Since we can choose 
arbitrarily the reference frames in
$\alpha$ and $\beta$, then the theory is invariant 
under arbitrary Poincar\'e trasformations $U(\alpha)$, 

\begin{eqnarray}
\Lambda(\alpha,\beta) 
&\mapsto& U(\alpha)\Lambda(\alpha,\beta)U^{-1}(\beta)\;\;\;\; \nonumber \\
x^{a}(\alpha)&\mapsto&
U^{a}_{b}(\alpha)x^{b}(\alpha)+U^{a}(\alpha)\;\;\;\;.
\label{invariance}
\end{eqnarray}

\noindent Now on we choose, following \cite{Ale}, to put the origin of each reference frame in the dual Voronoi vertices, then $x^{a}(\alpha)=0$ and the 
Poincar\'e group is restricted to the rotation matrices without translations. Then $\Lambda(\alpha,\beta)$ indicates only the rotation matrix.
Without imposing this sort of gauge fixing, it might be possible to study a metric-affine theory of gravity on lattice and this could be the subject of
a further research. See \cite{nonmetric} for more details on metric-affine theory of gravity case as well \cite{vergogna} and 
\cite{uomoe} for the relative consequences in the continuum case.         

\noindent The hinge $h$ is in one-to-one correspondence 
with its dual two-dimensional Voronoi 
plaquette that we still label by $h$. 
Now consider 
the following plaquette variable 

\begin{equation}
W_{\alpha}(h)=\Lambda(\alpha,\beta)
\Lambda(\beta,\delta^{h}_{1})...\Lambda(\delta^{h}_{k},\alpha)\;\;\;\; .
\label{ruota}
\end{equation}

\noindent As it has been shown \cite{hamber} \cite{thesis}, (\ref{ruota}) is a rotation
in the two-dimensional plane orthogonal to the $n-2$-dimensional hyperplane 
spanned by the hinge $h$. The rotation angle is the deficit angle (\ref{apex}).  

\noindent We consider $n-2$
linear independent edge-vectors of the hinge $h$ defined in the following way

\begin{equation}
E_{1}^{a}(\alpha)\equiv y^{a}_{1}(\alpha)-y^{a}_{n-1}(\alpha),...,
E_{n-2}^{a}(\alpha)\equiv y^{a}_{n-2}(\alpha)-y^{a}_{n-1}(\alpha), 
\label{vecto}
\end{equation}

\noindent and the antisymmetric tensor related to the oriented volume of $h$

\begin{equation}
{\mathcal V}^{(h)}{\;}^{ab}(\alpha)\equiv {1\over (n-2)!}
\epsilon^{ab}_{c_{1}...c_{n-2}}E_{1}^{c_1}(\alpha)...E_{n-2}^{c_{n-2}}(\alpha)\;\;\;\; .
\label{orio}
\end{equation}

\noindent At this point it seems natural to propose the following gravitational action \cite{Ale}

\begin{equation} 
I=-{1\over 2}\sum_{h}\Bigl( W_{\alpha}^{a_{1}a_{2}}{\;}^{(h)}
{\mathcal V}^{(h)}{\;}_{a_{1}a_{2}}(\alpha)\Bigr)
\label{haction}\;\;\;\;.
\end{equation}

\noindent Let $V(h)$ be the oriented volume of the hinge $h$.
The action (\ref{haction}) is equal \cite{Ale} to
 
\begin{equation}
I= \sum_{h}sin\;K(h)\;V(h)\;\;\;\; ,
\label{calc}
\end{equation}

\noindent which for small deficit angles $K(h)$ reduces to the Regge action (\ref{reggeact}).

\noindent The presence in the action of $sinK(h)$ instead of $K(h)$ is only a
lattice artifact \cite{Ale}. The useful regime of this theory is $sinK(h) \approx K(h)$. In fact  
Fr\"ohlich \cite{frohlich} has conjectured, as claimed also in \cite{Ale}, that this action converges to the Einstein-Hilbert action,
and J.Cheeger, W. Muller, R.Schrader in \cite{Cheeger} have proved that the Einstein-Regge action (\ref{reggeact})
converges to the Einstein-Hilbert action. They proved that the convergence of the action is in {\it sense of measure}.  
This applies when, roughly speaking, 
the number of the hinges of the simplicial manifold increases along with the incident number of the simplices at each hinge.
Then the triangulations will be finer and finer and the difference, in modulus, between the Einstein-Hilbert action on a manifold and
the Regge-Einstein one, on the triangulations of the same manifold, become smaller and smaller.         

\noindent Let's consider the $n-1$-dimensional face $f_{\alpha\beta}\equiv \{P_1,...,Q_{\alpha\beta}\}$ 
between the simplices $\alpha$ and $\beta$. The vertex coordinates of $f_{\alpha\beta}$ 
are $x_{i}^{a}(\alpha), a=1,...,n$. The following vector \cite{Ale}

\begin{eqnarray}
{b_{\alpha\beta}}{\;}_{a}(\alpha)&=&
\epsilon_{ab_{1}...b_{n-1}}(x_{1}}(\alpha)-x_{n}(\alpha))^{b_{1}}...
(x_{n-1}(\alpha)-x_{n}(\alpha))^{b_{n-1}\nonumber\\
&=&
\epsilon_{ab_{1}...b_{n-1}}E^{b_{1}}_{1}...E^{b_{n-1}}_{n-1}
\label{1tetra}
\end{eqnarray}

\noindent is normal to the face $f_{\alpha\beta}$, and it is assumed 
to point outward from the interior of the simplex $\alpha$. 
Simple considerations of linear algebra show 
that ${b_{\alpha\beta}}{\;}_{a}(\alpha)$ is an element of the inverse $n\times n$, 
matrix built using the components of $n$ linearly independent 
edge vectors of the simplex $\alpha$ multiplied by the determinant of the direct matrix (mathematically:
the adjoint elements, modulo the determinant, of the matrix of the n-independent vectors).  
The analogous vector ${b_{\beta\alpha}}{\;}_{a}(\beta)$ in the reference frame of $\beta$ is related 
to the previous one by 

\begin{equation}
{b_{\alpha\beta}}{\;}^{a}(\alpha)=
 \Lambda^{a}_{b}(\alpha,\beta){b_{\beta\alpha}}{\;}^{b}(\beta)\;\;\;\; .
\label{relatto}
\end{equation}

\noindent  It can be easily proved that (\ref{orio}) can be written as a bivector  

\begin{equation}
{\mathcal{V}}^{(h)}{\;}^{c_1c_2}(\alpha)\equiv
{1\over n!(n-2)!V(\alpha)}\Bigl({b_{\alpha\beta}}^{c_1}
(\alpha){b_{\alpha\delta^{h}_{k}}}^{c_2}(\alpha)
-{b_{\alpha\beta}}^{c_2}(\alpha){b_{\alpha,\delta^{h}_{k}}}^{c_1}(\alpha)\Bigr),
\label{arrota}
\end{equation}

\noindent where $V(\alpha)$ is the oriented volume of the simplex $\alpha$. 

\noindent We can consider the action as written on the dual Voronoi-complex of the 
original simplicial complex. The dual of the hinge $h$ is a two dimensional
plaquette whose vertices will be 
$\left\{\alpha,\beta,\delta^{h}_{1},...,\delta^{h}_{k}\right\}$. These vertices are the dual
of the $n$-simplices incident on the hinge $h$.

\noindent If, briefly, we indicate $\Lambda(\alpha,\beta)$ as $\Lambda_{\alpha\beta}$, and so on, the holonomy
matrix (\ref{ruota}) around the plaquette is

\begin{equation}
U^{h}_{\alpha \alpha}\equiv 
\Lambda_{\alpha\beta}\Lambda_{\beta\delta^{h}_{1}}
...\Lambda_{\delta^{h}_{k}\alpha}\;\;\;\;,
\label{holonomy}
\end{equation}

\noindent so that the action can be written in term of the bivectors ${b_{\alpha\beta}}{\;}^{a}(\alpha)$,

\begin{equation}
S\equiv -{1\over 2}\sum_{h}{\rm Tr}
\left(U^{h}_{\alpha\alpha}{\mathcal{V}}^{h}(\alpha)\right)
\label{daction}\;\;\;\;.
\end{equation}

\noindent As it has been remarked in \cite{Ale}, this sum i) does not
depend on the starting point of each two-dimensional plaquette chosen, and ii) it is a functional 
on the two dimensional plaquette of the Voronoi complex. 

\section{First-order set up}

Up to now we have dealt with a second order formalism of discrete General Relativity .
More precisely the connection matrices $\Lambda_{\alpha\beta}$ and the $b_{\alpha\beta}$
are both functions of the coordinates of the edges of the simplicial complex. 
Now we introduce a first
order formalism in which $\Lambda_{\alpha\beta}$ and
$b_{\alpha\beta}(\alpha)$ are independent variables. We set, only, the following constraints

\begin{equation}
b_{\alpha\beta}^{a}(\alpha)=
\Lambda_{\alpha\beta}{\;}^{a}_{b}b^{b}_{\beta\alpha}(\beta)\;\;,  
\label{duplex}
\end{equation}

\noindent which fix $n$ independent conditions for each face, not enogh to determine the ${n(n-1) \over 2}$ 
degrees of freedom of $\Lambda_{\alpha\beta}$. 

\noindent On the $b_{\alpha\beta}(\alpha)$ there is a further constraint
 since  the $n+1$ normals to the $n-1$-dimensional faces 
 of a $n$-simplex $\alpha$ are linearly dependent,

\begin{equation}
\sum_{\beta=1}^{n+1}b_{\alpha\beta}(\alpha)=0
\label{chiusura}\;\;\;\; .
\end{equation}

\noindent In this first-order formalism of General Relativity \cite{palatini} \cite{barrett} we will consider 
torsion-free, metric-compatible connection matrices more general than the Levi-Civita one defined by equation 
(\ref{connec}). (for a discussion on a possible weaking of the metricity condition see \cite{nonmetric}. Regge
Calculus with Torsion has been studied by I.T. Drummond \cite{torco})   
There is a technical problem we would like pointing out.  
The gravitational action 
in the form (\ref{daction}) could be
dependent on the starting simplex .
 So we need to define the following 
antisymmetric tensor on the plaquette $h$ 

\begin{eqnarray}
W^{(h)}_{c_1c_2}(\alpha) &\equiv & 
{1\over k_{h}+2} \Big({\mathcal{V}}^{(h)}(\alpha)+
\Lambda_{\alpha\beta}{\mathcal{V}}^{(h)}(\beta)\Lambda_{\beta\alpha}+... \nonumber\\
&+&\Lambda_{\alpha\beta}...
\Lambda_{\delta^{h}_{k-1}\delta^{h}_{k}}
{\mathcal{V}}^{(h)}(\delta^{h}_{k})\Lambda_{\delta^{h}_{k}\delta^{h}_{k-1}}...
\Lambda_{\beta\alpha} \Big)_{c_1c_2}
\label{media}\;\;\;\;.
\end{eqnarray}

\noindent The action then is 

\begin{equation}
S\equiv -{1\over 2}\sum_{h}{\rm Tr}
\left(U^{h}_{\alpha\alpha}W^{h}(\alpha)\right)
\label{caction}
\end{equation}

\noindent The action (\ref{caction}) coincides with the action (\ref{daction}) in the second-order formalism, and, 
since (in matrix notation)

\begin{equation}
W^{(h)}(\alpha)=\Lambda_{\alpha\beta}...\Lambda_{\delta_{i-1}\delta_{i}}
W^{(h)}(\delta_{i})\Lambda_{\delta_{i}\delta_{i-1}}...\Lambda_{\beta\alpha}
\label{1lastessa}\;\;\;\;.
\end{equation}

\noindent equation (\ref{caction}) is  independent of the starting simplex.  

\noindent Moreover the action (\ref{caction}) is invariant under the following set of 
transformations

\begin{eqnarray}
\Lambda_{\alpha\beta}&\mapsto&
{\Lambda}'_{\alpha\beta}=
O(\alpha)\Lambda_{\alpha\beta}O^{-1}(\beta)\nonumber\\
b_{\alpha\beta}(\alpha)&\mapsto&
b'_{\alpha\beta}(\alpha)=O(\alpha)b_{\alpha\beta}(\alpha)
\label{gauge}
\end{eqnarray}

\noindent where $O(\alpha)$ and $O(\beta)$ are two elements of $SO(n)$. This can be interpreted as the diffeomorphism
invariance of this discrete theory of Gravity. As we remarked before, without gauge-fixing, the affine group 
would have been the invariance group. This would have been the starting point to look for 
a description of Gravity, in the connection formalism, as a Yang-Mills theory of the affine group \cite{scassa} in
the discrete case. We highlighted this problem as a, possible, subject for a future investigation.

\begin{figure}

\epsfxsize=8.0truecm
\epsfysize=8.0truecm
\centerline{\hbox{\epsffile{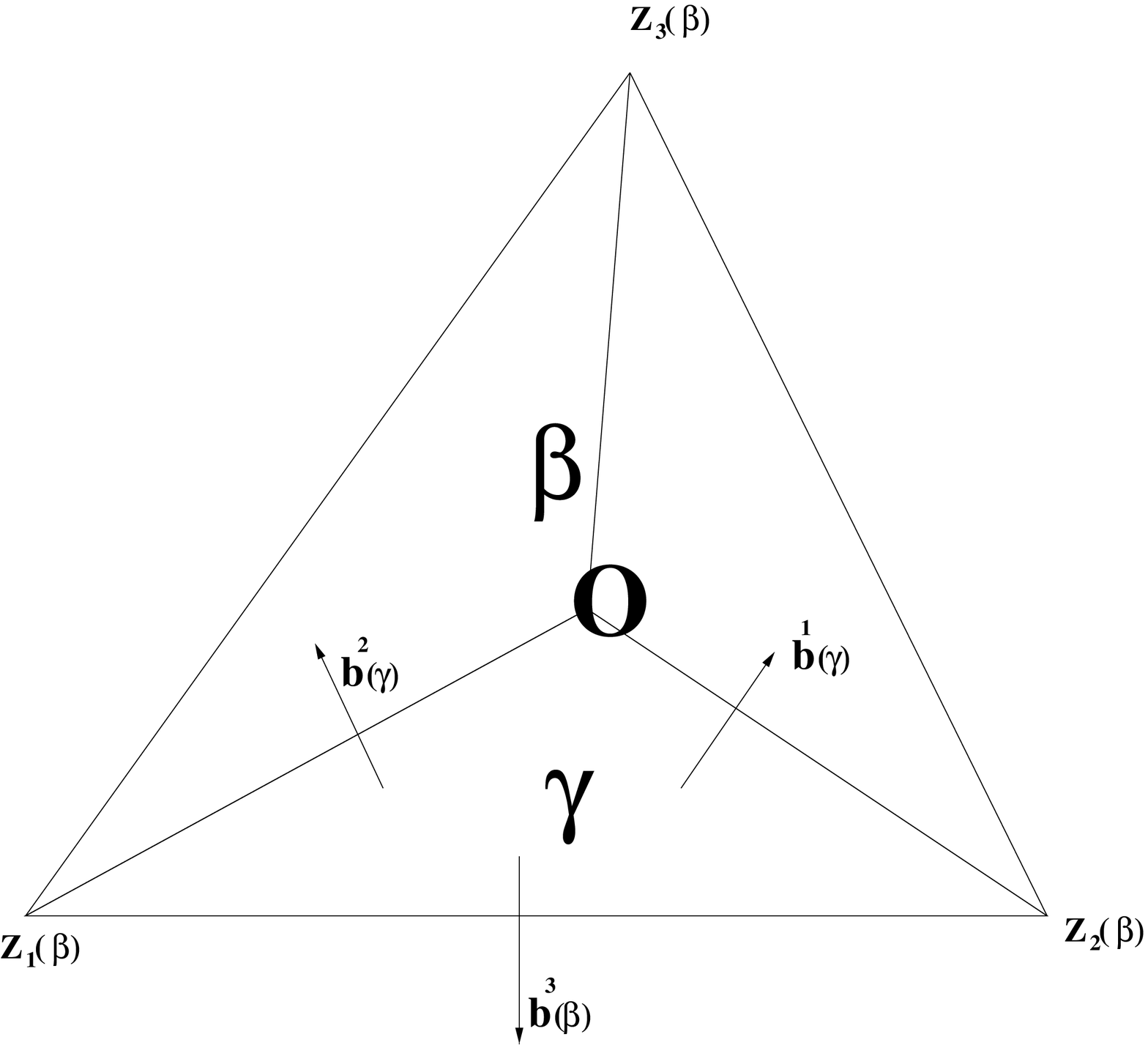}}}

\caption{\label{barycenter} Barycentric Coordinates}

\end{figure}

\section{$N$-bein on the simplicial complex}

It has been pointed out \cite{frohlich}\cite{Ale} that the $b_{\alpha \beta}$ is the $n$-bein
in the reference frame of each simplex. We are going to stress a feature which will be useful
in the next. It has been noticed in \cite{frohlich} that the way in which we have defined the 
coordinates, and, as consequence, the $b_{\alpha \beta}$ itself, 
does not take into account the symmetry of each simplex. 
A better definition consists in introducing affine or ``barycentric'' coordinates. 
Each point $P$ in the interior of a simplex $\beta$ can be considered as a barycenter of 
$n+1$ masses assigned at the vertices of the simplex (see reference \cite{Sorkin} and \cite{frohlich} for 
more details). In particular, if we choose in the $n$-simplex $\beta$ a 
reference frame in which the origin coincides 
with the geometric barycenter \cite{Ale}, the coordinates $z^{a}_{i}(\beta)$ 
of the vertices, in this reference frame, by the definition of the geometric barycenter, satisfy the condition

\begin{equation}
\sum_{i=0}^{n}z^{a}_{i}(\beta)=0         
\label{baric}
\end{equation}

\noindent which is analogous to (\ref{chiusura}). 
In barycentric coordinates \cite{Ale} we may write the 
$b_{\alpha \beta}$ as

\begin{equation}
b^{a}_{i}(\beta)={1\over (n-1)!}
\sum_{k\neq i}\epsilon^{a}_{a_1...a_{n-1}}\epsilon_{ki}^{j_1...j_{n-1}}
z^{a_1}_{j_1}(\beta)...z^{a_{n-1}}_{j_{n-1}}(\beta)
\label{circob}\;\;\;\; .
\end{equation}

\noindent The meaning of this formula can be understood, geometrically, 
in an easy way if we look at FIG. 2.
$b_{3}(\beta)$ is also normal to a face which belongs to $\gamma$. 
So it can be evaluated by solving (\ref{chiusura}), considered as an equation in $\gamma$, with respect to  
$b_{3}(\beta)$. Equation (\ref{circob}) does not give preference to the coordinates of any vertex. 
Thus the symmetry of the simplex is preserved as desired. The relation (\ref{circob}) can be inverted with respect 
to $z^{a}_{i}$ and, as in \cite{Ale}, we obtain

\begin{equation}
z^{a}_{i}(\beta)={n\over (n+1)!(n!V(\alpha))^{n-2}}
\sum_{k\neq i}\epsilon^{a}_{a_1...a_{n-1}}\epsilon_{ki}^{j_1...j_{n-1}}
b^{a_1}_{j_1}(\beta)...b^{a_{n-1}}_{j_{n-1}}(\beta)
\label{coppo}\;\;\;\; .
\end{equation}
 
\noindent It could be seen that (\ref{coppo}) is related to the dual barycentric base as 
explained in \cite{Sorkin} and \cite{frohlich}, but we do not explicitly discuss this link, 
since it is not relevant for the purpose of this article.

\section{First order field equations for small deficit angles}

As remarked in \cite{Ale}, in the second order 
formalism the action (\ref{daction}) is equivalent to the 
Regge action for small deficit angles $K(h)$. 
In our first order formalism we don't have angles $K(h)$. The variables
related to the deficit angles are the connection 
matrices $\Lambda_{\alpha\beta}$. Then we assume, by definition, 
that the {\it small deficit
angle approximation} in the first order formalism is the passage
from the group variables $\Lambda_{\alpha\beta}$ of 
SO(n) to the algebraic variables $\phi_{\alpha\beta}$ of so(n). 
As a consequence the 
connection matrices can be written in the form

\begin{equation}
\Lambda_{\alpha\beta}=I+\epsilon \; \phi_{\alpha\beta} + o(\epsilon)
\label{approx}
\end{equation}

\noindent In order to avoid the technical complications, which we shall discuss
in the next section, we now substitute
the constraint
$b_{\alpha\beta}(\alpha)=\Lambda_{\alpha\beta}b_{\beta\alpha}(\beta)$ into the  
action for each Voronoi edge $\alpha \beta$, which, in the current approximation, 
is 

\begin{equation}
b_{\alpha\beta}(\alpha)=
(I+\epsilon\;\phi_{\alpha\beta})b_{\beta\alpha}(\beta) + o(\epsilon)\;\;\;\;. 
\label{first}
\end{equation}

\noindent A straightforward explanation of the action up to the first order shows that

\begin{equation}
S=-{1\over 2}\epsilon\;
\sum_{h}Tr\left((\phi_{\alpha\beta}+\phi_{\beta\delta^{h}_{1}}...
+\phi_{\delta^{h}_{k}\alpha}){\;}^{0}W^{h}(\alpha)\right)+ o(\epsilon)
\label{expantion}
\end{equation}

\noindent where ${\;}^{0}W^{h}{\;}^{c_1c_2}$ is the $(2,0)$ antisymmetric tensor
(\ref{media}) at zeroth order, in the $\epsilon$-expansion. Here we have used 
the approximation (\ref{approx}). For each Voronoi-edge 
we have solved the constraint (\ref{duplex}) 
up to the first order (\ref{first}).

\noindent We also impose the requirement that the action is stationary under variation with respect 
to $\phi_{\alpha\beta}$, 

\begin{equation}
{\delta S \over \delta \phi_{\alpha\beta}{\;}_{c_1c_2}}=
\epsilon \sum_{h \in (\alpha\beta)}{\;}^{0}W^{h}{\;}^{c_1c_2}
+o(\epsilon)=0\;\;\;\;.
\label{1ord}
\end{equation}

\noindent These are the ${n(n-1)\over 2}$ implicit equations 
for $\phi_{\alpha\beta}$, which has ${n(n-1)\over 2}$
independent components. Since

\begin{eqnarray}
\fl \frac{\partial{\;}^{0}W^{h}{\;}^{c_1c_2}(\phi_{\rho\sigma},b_{i})}{\partial\phi_{\alpha\beta}{\;\;}^{r}_{s}}
=-{\epsilon\over n!(n-2)!V(\beta)}
\bigg[\Big(\delta^{c_1}_{r}b_{\beta\delta^{h}_{1}}^{c_1}b^{s}_{\alpha\beta}
-\delta^{c_1}_{s}b_{\beta\delta^{h}_{1}}^{c_2}b^{r}_{\alpha\beta}\Big)\nonumber \\
-\Big(\delta^{c_2}_{r}b_{\beta\delta^{h}_{1}}^{c_1}b^{s}_{\alpha\beta}-
\delta^{c_2}_{s}b_{\beta\delta^{h}_{1}}^{c_1}b^{r}_{\alpha\beta}\Bigr)\bigg],
\label{condition}
\end{eqnarray}
   
\noindent the determinant of this matrix is non zero 
(notice that matrix (\ref{condition}) has dimension ${n(n-1)\over 2}$). By inverse function 
theorem (also labeled Dini's theorem), 
locally, 
we can invert equation (\ref{1ord}), so that, again locally, there is one solution  
giving $\phi_{\alpha\beta}$ as a function of the $n$-bein $b_{i}$. 
We do not know anything about the uniqueness of this solution globally. 
We are, in some sense, in a 
situation analogous to that described in \cite{barrett}. Barrett found, locally, the uniqueness of the solution, 
but has showed that this result is not valid globally. 
Locally the unique solution can be found by using the $Levi\; Civita-Regge$ connection. We determine the
matrix connection by equations (\ref{connec}), in which the translation vector is a known quantity,
once we put the origins of the refernce frames in the barycenter of the simplices. The equations
(\ref{connec}) determine the connection matrices as functions of the coordinates of the vertices. In 
barycentric coordinates these are in one-to-one relation with the $b_{i}$. Then we determine 
the connection matrices as functions of the $b_{i}$.   
The identity

\begin{equation} 
\sum_{h\in (\alpha\beta)} 
{\mathcal{V}}^{h}{\;}^{c_1c_2}(\alpha)
=0 \;\;\;\;,
\label{identity}
\end{equation}
 
\noindent can be easily proved if we express 
${\mathcal{V}}^{(h)}{\;}^{c_1c_2}(\alpha)$
as in (\ref{arrota}), and use the constraint 
$\sum_{\beta=1}^{n+1}b^{a}_{\alpha\beta}(\alpha)=0$.
The statement that this is the Levi Civita-Regge
connection implies  

\begin{equation}
{\mathcal{V}}^{(h)}(\alpha)
=\Lambda_{\alpha\beta}{\mathcal{V}}^{(h)}(\beta)\Lambda_{\beta\alpha}
=...=\Lambda_{\alpha\beta}...\Lambda_{\delta^{h}_{k-1}\delta^{h}_{k}}
{\mathcal{V}}^{(h)}(\delta^{h}_{k})\Lambda_{\delta^{h}_{k}\delta^{h}_{k-1}}...
\Lambda_{\beta\alpha} 
\label{equo0}\;\;\;\;,
\end{equation}

\noindent so that to the zero order 

\begin{equation}
{\mathcal{V}}^{(h)}(\alpha)={\mathcal{V}}^{(h)}(\beta) +O(\epsilon)=...=
{\mathcal{V}}^{(h)}(\delta_{k}^{h})+O(\epsilon)
\label{requo}\;\;\;\;.
\end{equation}

\noindent These facts imply that

\begin{equation}
\epsilon\sum_{h \in (\alpha\beta)}{\;}^{0}W^{h}{\;}^{c_1c_2}
=\sum_{h \in (\alpha\beta)}
\left(\epsilon{\mathcal{V}}^{h}{\;}^{c_1c_2}(\alpha) + \epsilon O(\epsilon)\right)      
=0+O(\epsilon^{2})
\label{soddisf}\;\;\;\;,
\end{equation}

\noindent that is to say the Levi Civita-Regge connection is the solution of our first-order 
equations for the connection matrices in the limit of small deficit angles.

\section{First order field equation: the general case}

In the previous section we have seen that, in the case of small deficit angles,  
Regge Calculus is the solution of
the first order field equations. 

\noindent Now  we are going to deal with the general problem. We would like to derive 
the equation of motion by varying the action with respect to $\Lambda_{\alpha\beta}$
and $b_{\alpha\beta}$. First we have to take in account the constraints
(\ref{duplex}) and (\ref{chiusura}).  
So, in order to perform independent
variations of $\Lambda_{\alpha\beta}$ and $b_{\alpha\beta}$, it is necessary to put
 Lagrange multipliers in the action. Then the action, in the Palatini first order, is

\begin{eqnarray}
S&\equiv& -{1\over 2}\sum_{h}Tr\left(U^{h}_{\alpha\alpha}W^{h}(\alpha)\right)+
\sum_{(\alpha\beta)}\lambda_{\alpha\beta}\left(b_{\alpha\beta}(\alpha) -
\Lambda_{\alpha\beta}b_{\beta\alpha}(\beta)\right) \nonumber \\
&+& \sum_{(\alpha\beta)}Tr\left({\tilde{\lambda}}^{(\alpha\beta)}
\bigl(\Lambda_{\alpha\beta}\Lambda^{T}_{\alpha\beta}-I\bigr)\right)
+\sum_{\alpha}\mu(\alpha)\left(\sum_{\beta =1}^{n+1}
b_{\alpha\beta}(\alpha)\right) 
\label{1action}\;\;\;\;,
\end{eqnarray}

\noindent where $\lambda^{(\alpha\beta)}$ and $\mu(\alpha)$ are $n$-dimensional vectors 
and Lagrange
multipliers respectively, and ${\tilde{\lambda}}^{(\alpha\beta)}$ is an $n \times n$ matrix. The
constraint $\Lambda_{\alpha\beta}\Lambda^{T}_{\alpha\beta}-I$ is introduced to
restrict the variation of $\Lambda_{\alpha\beta}$ on the group $SO(n)$. 

\noindent We introduce the following lemma:

\noindent {\bf Lemma}: {\it The action, in matrix notation, 

\begin{equation}
S'\equiv Tr(\Lambda A) + Tr\left(\lambda(\Lambda\Lambda^{T}-I)\right)
\label{lemma}
\end{equation}

\noindent gives the equation of motion, (if we assume that the variation with 
respect to $\Lambda$ is stationary})

\begin{equation}
(\Lambda A)= (\Lambda A)^{T}
\label{reslem}\;\;\;\; .
\end{equation}

\noindent {\bf Proof}: If we consider the variation 
of the action, using the property that $Tr(M)=Tr(M^{T})$ and the action is stationary respect to 
$\Lambda$ and $\Lambda^{T}$ we have

\begin{eqnarray}
{0 \equiv {\delta S'\over \delta\Lambda}}&=&A+
\Lambda^{T}\lambda\;\;\;\;\;,\nonumber\\
{0 \equiv {\delta S'\over\delta\Lambda^{T}}}&=& A^{T}+
\lambda\Lambda
\label{equo}\;\;\;\;.
\end{eqnarray}

\noindent Multiplying the first equation for $\Lambda$ on the left, and the second
for $\Lambda^{T}$ on the right, and subtracting term by term these two equations, we have (\ref{reslem}).

\noindent Applying the lemma to the action (\ref{1action}) for the 
variation with respect to $\Lambda_{\alpha\beta}$, we obtain the
field equations

\begin{equation}
\fl \sum_{h\in (\alpha\beta)}(U^{h}_{\alpha\alpha}W^{h}(\alpha))_{ij}
-\lambda_{\alpha\beta}{\;}_{i}b_{\alpha\beta}(\alpha){\;}_{j}=
\sum_{h\in (\alpha\beta)}(U^{h}_{\alpha\alpha}W^{h}(\alpha))^{T}_{ij}
-\lambda_{\alpha\beta}{\;}_{j}b_{\alpha\beta}(\alpha){\;}_{i}
\label{variola}\;\;\;\;.
\end{equation}

\noindent The next step will be to determine the field equations for the variations
of $b_{\alpha\beta}$. For this it is necessary to determine the 
quantity ${\partial V(\alpha) \over \partial b^{a}_{\alpha\beta}}$.
As remarked in \cite{Ale}, we have the following identity

\begin{equation}
\left(V(\alpha)\right)^{n-1}={1\over {n!}^{n}}
\epsilon_{a_1...a_{n}}\epsilon^{j_1...j_{n} j}
b^{a_1}_{j_1}(\beta)...b^{a_{n}}_{j_{n}}(\beta)
\label{vol}\;\;\;\;.
\end{equation}

\noindent We can write this formula in a way which is not dependent
on the chosen index $j$.Equivalently equation (\ref{vol}) can be written as

\begin{equation}
\left(V(\alpha)\right)^{n-1}={1\over {n!}^{n}}
\frac{1}{n+1}\sum_{j=1}^{n+1}\epsilon_{a_1...a_{n}}\epsilon^{j_1...j_{n} j}
b^{a_1}_{j_1}(\beta)...b^{a_{n}}_{j_{n}}(\beta)
\label{medvol}\;\;\;\; ,
\end{equation}

\noindent so that it is clearly independent of any index. Thus, we have

\begin{equation}
V(\alpha)= \root{n-1} \of {\left({1\over {n!}^{n}}
\frac{1}{n+1}\sum_{j=1}^{n+1}\epsilon_{a_1...a_{n}}\epsilon^{j_1...j_{n} j}
b^{a_1}_{j_1}(\alpha)...b^{a_{n}}_{j_{n}}(\alpha)\right)}
\label{radix}\;\;\;\; .
\end{equation}

\noindent Evaluating 
its derivative with respect to $b^{a}_{i}(\alpha)$ we obtain

\begin{equation}
{\partial \left(1\over V(\alpha))\right)\over \partial b^{a}_{i}(\alpha)}=
-{1\over {n!(n-1)V^{2}(\alpha)}}z^{i}_{a}(\alpha)
\label{derivo}\;\;\;\; .
\end{equation}

\noindent Now we are ready to derive the field equations for the variations of the 
$b_{\alpha\beta}$. So let's consider the action (\ref{1action}). It is straightforward to 
write it as 

\begin{eqnarray}
\fl S\equiv -{1\over 4}\sum_{h}Tr\left(U^{h}_{\alpha\alpha}W^{h}(\alpha)
+{U}^{T}{\;}^{h}_{\alpha\alpha}{W}^{T}{\;}^{h}(\alpha)\right)+
\sum_{(\alpha\beta)}\lambda^{(\alpha\beta)}\left(b_{\alpha\beta}(\alpha) -
\Lambda_{\alpha\beta}b_{\beta\alpha}(\beta)\right) \nonumber \\
+\sum_{(\alpha\beta)}Tr\left({\tilde{\lambda}}^{(\alpha\beta)}
\bigl(\Lambda_{\alpha\beta}\Lambda^{T}_{\alpha\beta}-I\bigr)\right) +
\sum_{\alpha}\mu(\alpha)\left(\sum_{\beta =1}^{n+1}
b_{\alpha\beta}(\alpha)\right) 
\label{2action}\;\;\;\; .
\end{eqnarray}

\noindent Let's define 

\begin{equation}
U^{h}_{\alpha\alpha}{\;}^{ij}-U^{h}_{\alpha\alpha}{\;}^{ji}\equiv
\Omega^{h}_{\alpha\alpha}{\;}^{ij}\;\;\;\; ,
\end{equation}

\noindent the variation with respect to $b^{i}_{\alpha\beta}(\alpha)$ is 

\begin{eqnarray}
\fl {\partial S \over \partial b^{i}_{\alpha\beta}(\alpha)}\equiv
\sum_{h\in (\alpha\beta)}{1\over 4 n!(n-2)! V(\alpha)}b_{\alpha\delta_{k}^{h}}^{j}
\left(\Omega^{h}_{\alpha\alpha}{\;}_{ij}
-{2\over n!(n-1)V(\alpha)}\Omega^{h}_{\alpha\alpha}{\;}_{kj}b_{\alpha\beta}^{k}(\alpha)
  z^{\alpha\beta}_{i}(\alpha)\right)\nonumber\\
  +\lambda_{i}{\;}_{\alpha\beta}-\mu_{i}(\alpha)=0
\label{moto}\;\;\;\;.
\end{eqnarray}

\noindent The field equations (\ref{moto}) and (\ref{variola}) 
are the analogues of the first order Einstein 
equations in the $n$-bein formalism (see \cite{myself} for a strict analogy).  
The connection matrices, $\Lambda_{\alpha\beta}$, the $n$-bein, $b^{i}_{\alpha\beta}(\alpha)$, and the 
Lagrange multipliers $\lambda_{\alpha\beta}^{i}$ and $\mu^{i}(\alpha)$ must 
be determined by solving the equations (\ref{moto}) and (\ref{variola}) and the constraints (\ref{duplex}), 
(\ref{chiusura}) as well the orthogonal condition of the connection matrices. 

\noindent A nice feature of these equations is that if 
$\Bigl\{\Lambda_{\alpha\beta}, b^{i}_{a}(\beta),\lambda_{i}{\;}_{\alpha\beta},\mu_{i}(\alpha)\Bigr\}$ is solution 
of the equations of motion, the map 

\begin{eqnarray}
\Lambda_{\alpha\beta} &\mapsto& 
\Lambda_{\alpha\beta}\;\;\;\;,\nonumber\\
b^{i}_{a}(\beta) &\mapsto& \omega b^{i}_{a}(\beta)
\;\;\;\; ,\nonumber\\
\lambda_{i}{\;}_{\alpha\beta} &\mapsto& 
\omega^{-{1\over n-1}} \lambda_{i}{\;}_{\alpha\beta}\nonumber\\
\mu_{i}(\alpha) &\mapsto& 
\omega^{-{1\over n-1}} \mu_{i}(\alpha)
\label{lagracon}\;\;\;\;.
\end{eqnarray}

\noindent will provide an other solution, as it is easily to verify, of the discrete Einstein equations. This map, if we restrict to the b's only, 
shows that the discrete Einstein equations have a conformal symmetry, anyway in the {\it solutions} not in the {\it action}. This is the same
feature of the Regge equations \cite{regge}.

\section{The measure}

In this section we shall discuss the quantum measure to associate with the
previous classical action. The action \cite{Ale} is invariant under the 
action of the group $SO(n)$, that is the gauge group (\ref{gauge}).

\noindent We will use the following notation:

\begin{equation}
\mu(b_{\alpha\beta}(\alpha))\equiv 
db^{1}_{\alpha\beta}(\alpha)...db^{n}_{\alpha\beta}(\alpha)
\label{bmes}\;\;\;\;,
\end{equation}

\noindent and let

\begin{equation}
\mu(\Lambda_{\alpha\beta})
\label{Haar}
\end{equation}

\noindent be  the Haar measure on $SO(n)$. The partition function for this theory is:

\begin{equation}
\fl Z=\int e^{{1\over 2}\sum_{h}Tr\left(U^{h}_{\alpha\alpha}W^{h}(\alpha)\right)}
\prod_{\alpha}\delta(\sum_{\beta=1}^{n+1}b_{\alpha\beta}(\alpha))
\prod_{\alpha\beta}
\delta(b_{\alpha\beta}(\alpha)-\Lambda_{\alpha\beta}b_{\beta\alpha(\beta)})
\mu(\Lambda_{\alpha\beta})\mu(b_{\alpha\beta})\;\;\;\;.
\label{part}
\end{equation}

\noindent Here the product $\prod_{\alpha}$ is a product over all vertices 
of the dual complex as well the  product $\prod_{\alpha\beta}$ 
over all the Voronoi links which has $\alpha$ as one of his vertices. 
It is straightforward to see that the measure 
is invariant under the gauge transformation (\ref{gauge}). 
In fact if we perform the gauge transformation
(\ref{gauge}), the modifications to the measure are:

\begin{equation}
\delta(\sum_{\beta=1}^{n+1}b'_{\alpha\beta}(\alpha))
\delta(b'_{\alpha\beta}(\alpha)-{\Lambda}'_{\alpha\beta}b'_{\beta\alpha(\beta)})
\mu({\Lambda}'_{\alpha\beta})\mu(b'_{\alpha\beta}) 
\label{modifico}\;\;\;\; .
\end{equation}

\noindent But $\mu(b'_{\alpha\beta})$ is equal to 
$det\Bigl(O(\alpha)\Bigr)\mu(b_{\alpha\beta})$, and reduces to
$\mu(b_{\alpha\beta})$ as well. The Haar measure of $SO(n)$ is right and left 
invariant. So $\mu({\Lambda}'_{\alpha\beta})=\mu({\Lambda}_{\alpha\beta})$.
Again, by using  equations (\ref{gauge}) the two deltas of
(\ref{modifico}) can be written in the following form:

\begin{eqnarray}
\delta\Bigl(O(\alpha)(\sum_{\beta=1}^{n+1}b_{\alpha\beta}(\alpha))\Bigr)\;\;\;\;,\nonumber\\
\delta\Bigl(O(\alpha)
(b_{\alpha\beta}(\alpha)-{\Lambda}_{\alpha\beta}b_{\beta\alpha(\beta)})\Bigr)
\label{soso}\;\;\;\;.
\end{eqnarray}

\noindent Then the properties of the delta function along with the last considerations prove  
that equation (\ref{modifico})
can be written as 

\begin{equation}
\delta(\sum_{\beta=1}^{n+1}b_{\alpha\beta}(\alpha))
\delta(b_{\alpha\beta}(\alpha)-{\Lambda}_{\alpha\beta}b_{\beta\alpha(\beta)})
\mu({\Lambda}_{\alpha\beta})\mu(b_{\alpha\beta}) 
\label{rivedo}\;\;\;\; ,
\end{equation}

\noindent This establishes the invariance of the measure under gauge trasformations. 

\noindent Challenged by \cite{menotti1} we can argue that the metric structure we 
are considering is not as peculiar as the metric written as a function of the edge-lengths 
in Regge Calculus. 
In first-order formalism we neither have transition functions 
which depend on the metric structure, the 
$n$-bein in our case, nor do we sum over a metric that is gauge-fixed. 
From the form of the constraint equations, 
those in the argument of the delta functions, and from the calculations we have 
performed, we expect that this measure, once we have integrated over the deltas, is really highly
not local as  has been discovered for Regge Calculus.          

\section{Coupling with matter}

\noindent In the continuum theory on Riemannian manifolds with torsion-free 
connection the coupling with fermionic matter (for the same case in presence
of torsion see \cite{uomoe}) is given 
by the lagrangian density

\begin{equation}
{\mathcal L}\equiv \frac{i}{2}\left({\bar \psi}e^{\mu}_{a}\gamma^{a}{\nabla}_{\mu}\psi
- e^{\mu}_{a}{\nabla}_{\mu}{\bar \psi}\gamma^{a}\psi\right)
-m\psi {\bar \psi}
\label{lagra}\;\;\;\;.
\end{equation}

\noindent The  $\gamma^{a},\; a=1,...n$ are the Dirac-matrices satisfying the Clifford
algebra 

\begin{equation}
\gamma^{a}\gamma^{b}+\gamma^{b}\gamma^{a}=2\delta^{a\;b}
\label{gamma}\;\;\;\;,
\end{equation}

\noindent whereas $\psi$ is  the n-dimensional Dirac spinor field 
(${\bar \psi}\equiv {\psi}^{\dagger}{\gamma}^{1}$),${\nabla}_{\mu}$ the
covariant derivative, and $e_{a}^{\mu}$ the n-beins
on the tangent space of the Riemannian manifold $(M,g)$, 
where the lagrangian density is 
defined  (\ref{lagra}), such that

\begin{equation}
g^{\mu\nu}(x)={e^{\mu}_{a}}(x){e^{\nu}_{b}}(x)\delta^{ab}
\label{tetra}\;\;\;\; .
\end{equation}
 
\noindent We are assuming that the Riemannian manifold $(M,g)$ in question has a
spin structure, that is its second Stiefel-Whitney class is zero.

\noindent Now we have all the ingredients to define the coupling of gravity with 
fermionic matter on the lattice in analogy with the continuum case. 
Let $2\nu=n$ or $n=2\nu+1$ (depending on whether 
$n$ is even or odd) and consider the $2^{\nu}$-dimensional representation
of the two-fold covering group of $SO(n)$\cite{frohlich}. So, instead of considering the
connection matrices $\Lambda_{\alpha\beta}$, we will deal with the
$2^{\nu}\times 2^{\nu}$ connection matrices $D_{\alpha\beta}$ such that

\begin{equation}
D_{\alpha\beta}{\gamma}^{a}D_{\alpha\beta}^{-1}
=(\Lambda_{\alpha\beta})^{a}_{b}{\gamma}^{b}
\label{2-fold}\;\;\;\;.
\end{equation}

\noindent Given $D_{\alpha\beta}$ we can determine
$\Lambda_{\alpha\beta}$. Furthermore if we know $\Lambda_{\alpha\beta}$, we can determine
$D_{\alpha\beta}$ up to a sign. In particular, from 
(\ref{2-fold}), we can write $\Lambda_{\alpha\beta}$ as \cite{frohlich}

\begin{equation}
(\Lambda_{\alpha\beta})_{ab}={1\over
2^{\nu}}Tr(\gamma_{a}D_{\alpha\beta}\gamma_{b}D_{\alpha\beta}^{-1})
\label{questa}
\end{equation}

\noindent In the discrete theory we assume that the spinor field is a $2^{\nu}$ complex 
vector defined at each vertex of the dual Voronoi complex, that is to say 
a map that to each vertex $\alpha$ associates the $2^{\nu}$ complex 
vector $\psi(\alpha)$.

\noindent In order to define the covariant derivative on a lattice we have to 
derive the distance $|\alpha\beta|$ between the two neighboring circumcenters
in $\alpha$ and $\beta$. Our reasoning concerning the baricenters can be extended 
to the circumcenters too. The distance $\triangle h_{1}$ of the circumcenter
in $\alpha$ from the face $\alpha\beta$ can be determined by calculating
the volume of the $n$-dimensional simplex obtained by joining the circumcenter
with the $n$-vertices of the face $\alpha\beta$, and dividing it by $n$ and by
the volume of the face itself. So we have

\begin{equation}
\triangle h_{1}={1\over
n^{3}}{\sum_{i=1}^{n}b^{a}_{\alpha\beta}(\alpha)z_{a}^{i}(\alpha) \over
|b_{\alpha\beta}(\alpha)|}
\label{dist}\;\;\;\;,
\end{equation}

\noindent in which the $z^{a}_{i}$ are the circumcentric coordinates of the vertices of the
face $\alpha\beta$ and $|b_{\alpha\beta}(\alpha)|$ the module of the 
$b_{\alpha\beta}(\alpha)$ written as function of the $z^{a}_{i}$, as in (\ref{circob}).

\noindent In the same manner we have

\begin{equation}
\triangle h_{2}={1\over
n^{3}}{\sum_{i=1}^{n}b^{a}_{\beta\alpha}(\beta)z_{a}^{i}(\beta) \over
|b_{\beta\alpha}(\beta)|}\;\;\;\;,
\label{dist2}
\end{equation}

\noindent Finally, we have $|\alpha\beta|=\triangle h_{1} + \triangle h_{2}$.

\noindent We are ready to define the covariant derivative
$(\nabla_{\mu}\psi)(\alpha)$ on a lattice

\begin{equation}
(\nabla_{\mu}\psi)(\alpha)\equiv
{ D_{\alpha\beta}\psi(\beta) - \psi(\alpha)\over |\alpha\beta|}
\label{covariant}\;\;\;\;.
\end{equation}

\noindent So far the discrete version of the the action for the coupling between gravity
and fermionic matter can be written in the form (see also \cite{ren})

\begin{eqnarray}
\fl S_{F} \equiv \sum_{\alpha}\Biggl(\sum_{(\alpha\beta), \beta=1,...,n+1}
{i\over 2|\alpha\beta|}\biggl({\bar \psi}(\alpha)b^{a}_{\alpha\beta}\gamma_{a}
D_{\alpha\beta}\psi(\beta)\nonumber \\  
- D_{\alpha\beta}{\bar\psi}(\beta)b^{a}_{\alpha\beta}\gamma_{a}\psi(\alpha)\biggr)-m V(\alpha){\bar{\psi(\alpha)}}\psi(\alpha)\Biggr)\;\;\;\;.
\label{fermion}
\end{eqnarray}

\noindent Then the quantum measure, which also includes fermionic matter, can be written 
as

\begin{eqnarray}
\fl Z=\int e^{-(S+S_{F})}
\prod_{\alpha}\mu(\psi(\alpha))\mu({\bar
\psi}(\alpha))\delta(\sum_{\beta=1}^{n+1}b_{\alpha\beta}(\alpha))\nonumber\\
\times\prod_{\alpha\beta}
\delta(b_{\alpha\beta}(\alpha)-\Lambda_{\alpha\beta}b_{\beta\alpha(\beta)})
\mu(D_{\alpha\beta})\mu(b_{\alpha\beta})
\label{minestra}
\end{eqnarray}

\noindent where $S$ is the action for pure gravity (\ref{caction}),  
$\mu(D_{\alpha\beta})$ the Haar measure on the two-fold covering group 
of $SO(n)$, while $\mu(\psi(\alpha))=d\psi(\alpha)$ is the standard measure 
on $C^{2^{\nu}}$. 

\section{Conclusions}

In this paper we have studied a discrete 
theory of gravity in its first-order formalism. Following an earlier 
previous paper \cite{Ale},  we have chosen 
an orthonormal reference frame in each simplex and have defined a connection matrix 
as a transformation matrix between the reference
frames of two $n$-dimensional simplices that share a common $n-1$-face, considered as 
two distinct reference frames at the same point. This defines a Levi Civita-Regge connection. 
These matrices allow us to write an holonomy matrix around each hinge
$h$. We define the action as the sum over the hinges of the traces of
the holonomy matrices multiplied by the oriented volumes of the hinges. This action can be
written on the dual Voronoi complex of the original simplicial complex. We can express
the action as a function of the connection matrices 
and of the vectors $b_{\alpha\beta}$ which are normal to the faces of the simplices and 
whose modulo is proportional to the volume of the faces themself.

\noindent The action is very similar to the Wilson action of lattice gauge theory. 
Here the $b_{\alpha\beta}$ have the same role as the $n$-bein in the 
continuum theory. Moreover the action is locally invariant under the
action of the Poincar\'e group.

\noindent On the dual Voronoi metric complex the first-order formalism is implemented by considering 
the dynamical variables $\Lambda_{\alpha\beta}$ and 
$b_{\alpha\beta}$ independent. It is shown that in the limit of small deficit angles the 
{\it Levi-Civita or Regge connection} is a solution of the equations of motion. As in Barrett \cite{barrett}, the solution
is unique locally, by Dini's theorem. However we do not know what is going to happen globally.  

\noindent The general equations for $\Lambda_{\alpha\beta}$ and $b_{\alpha\beta}$ are derived
by using   Lagrange multipliers, in order to take in account the 
constraints of the first order formalism. 

\noindent A quantum measure and the relative partition function has been defined. They are locally invariant 
under the action of $SO(n)$. We have introduced a coupling of gravity 
with fermionic matter on the Voronoi complex as well. This coupling seems as natural as 
for the continuum theory. In particular it seems the Voronoi complex allows us  to introduce the coupling
with matter, avoiding all the troubles we had in Regge Calculus. We hope that this first-order 
formalism might be useful for numerical simulations. In particular its first-order
character, numerically, could have more advantages than the usual (Regge) second order formulation 
in implementing the evolution of equations of motion.

\ack
I would like to thank Alessandro D'Adda for the early and determinant 
collaboration on this topic and Ruth Williams 
for discussions and helpful comments on this work. This research 
has been, partially, supported by {\it Fondazione della Riccia}
while I was at U.C. Irvine. I would like to express my gratitude
to Fr. Bill Stoeger S.J. for directing, encouraging, and advising me to complete this work, 
as well the moral support of Fr. Secondo Bongiovanni S.J., Fr. Gianluigi Brena S.J. and Fr. Giuseppe Pirola S.J. 
It is a pleasure to thank Fr. George Coyne S.J. and Fr. Francesco Tata S.J. for contributions
to the last stages of this research.

\end{document}